\documentstyle[epsf]{mn}  

\title[The colour--magnitude relation]{ 
{\vglue -47pt
\rightline{\LARGE\bf OU-TAP 100}
\vglue 26pt}
What determines the slope of the colour--magnitude relation of
elliptical galaxies in the hierarchical clustering scenario?}
\author[M. Nagashima and N. Gouda]{Masahiro Nagashima$^{1}$
  \thanks{masa@vega.ess.sci.osaka-u.ac.jp}
and Naoteru Gouda$^2$\thanks{naoteru.gouda@nao.ac.jp}\\
$^1$Department of Earth and Space Science, Graduate School of Science,
Osaka University, Toyonaka, Osaka 560-0043, Japan\\
$^2$National Astronomical Observatory, Mitaka, Tokyo 181-8588, Japan\\
}

\begin{document}   

\maketitle   

\begin{abstract}   
  We examined the physical origin of the colour--magnitude relation of
  elliptical galaxies in clusters of galaxies in the hierarchical
  clustering scenario by using a semi-analytic model of galaxy
  formation.  The semi-analytic model includes some physical processes
  connected with the galaxy formation, such as merging histories of
  dark haloes, gas cooling, star formation, supernova feedback,
  mergers of galaxies, and so on.  Therefore it is easier to
  investigate how the galaxy formation and evolution depend on these
  physical processes by using the semi-analytic model rather than
  other numerical methods such as N-body and hydrodynamical
  simulations.  Since the galaxy formation process is very
  complicated, analysis by using the semi-analytic model is effective.
  We particularly investigate the effects of the star formation
  process, the supernova feedback and the UV background on the
  formation and evolution of elliptical galaxies.
  
  It is suggested by observation that the colour--magnitude relation
  reflects the metallicity--luminosity relation of galaxies, and that
  for reproducing the colour--magnitude relation it is important to
  suppress the chemical enrichment of stars in faint galaxies.  For
  suppressing the chemical enrichment, we propose some solutions.  One
  is assuming a strong supernova feedback.  Another is introducing the
  effect of the UV background radiation.  These solutions degenerate,
  but they may be distinguished by studying other properties of
  galaxies, like the luminosity function, colour distribution, etc.
  
  Moreover, we find that the supernova feedback model in the starburst
  which make the bulge component is important when the star formation
  time-scale in disc is longer than the age of the Universe.  On the
  other hand, when the star formation time-scale is short, the
  resulting colour--magnitude relation does not depend on such a
  starburst model because most of stars are formed not by the burst
  but by the ordinary star formation process in disc.
\end{abstract}   
   

\section{INTRODUCTION}   
It is well known that elliptical galaxies in clusters of galaxies have
the tight correlation between their colours and magnitudes (so-called
`colour--magnitude relation'; hereafter CMR).  For example, the rms
scatter about the mean CMR is typically $\sim$ 0.04 mag in Virgo and
Coma clusters of galaxies.  It is a comparable size to observational
errors (Bower, Lucey \& Ellis 1992).  Because it is believed that this
tight relation reflects the formation and evolution processes of
elliptical galaxies, many people have studied this relationship in
order to understand the galaxy formation process.

The traditional scenario of the formation of ellipticals is that a
monolithic protogalactic cloud collapses and then forms stars for a
short time-scale until blowing the galactic wind (Larson 1974; Arimoto
\& Yoshii 1986, 1987).  In this framework, the {\it conventional}
interpretation of the CMR of ellipticals is considered as follows.
Colour of each galaxy reflects its mean stellar age and metallicity,
while luminosity of each galaxy reflects its mass.  Therefore, there
are two scenarios to make the difference of colours as follows.  One
is that redder luminous galaxies are formed at earlier stages of
cosmological structure formation, and bluer faint galaxies are formed
recently.  Another scenario is that elapsed time of redder galaxies
between the beginning of star formation and the epoch of galactic wind
is longer than that of bluer faint galaxies, then more metal-rich
stars are formed in luminous galaxies.  Therefore the luminous
galaxies have redder colour.  These two effects, age and metallicity,
for reddening the luminous galaxies degenerate (so-called
`age--metallicity degeneracy'; Worthey 1996 and Arimoto 1996).  This
degeneracy is solved in the collapse/wind model by Kodama \& Arimoto
(1997) by comparing the CMRs of high redshift cluster galaxies with
those of theoretical models.  If the CMR reflects the {\it age
  sequence}, young faint galaxies should be very blue at high
redshift, and the slope of the CMRs evolves rapidly.  However, the
difference between the slopes of the CMRs at the present epoch and
those at high redshift is small compared to the theoretical prediction
in the age sequence.  This small difference is in agreement with the
prediction in the {\it metallicity sequence}.  Moreover, they
suggested that age of ellipticals is longer than $\sim 10$ Gyr.  Thus,
they found that the CMR reflects the metallicity sequence.

In contrast to such a collapse/wind model, recent developments on both
theory and observation of the cosmological structure formation are
revealing that objects such as galaxies and clusters of galaxies are
formed through hierarchical clustering of smaller objects.  Kauffmann
\& Charlot (1998; hereafter KC) applied their semi-analytic model to
the problem of the CMR, and reproduced the observational properties of
the CMRs, when the model includes the chemical evolution process and
strong feedback to interstellar media by supernovae.  Moreover, they
found a tight luminosity--metallicity relation.

KC interpreted why the CMR is reproduced when the feedback process is
strong as follows.  They considered that bright ellipticals must be
formed by mergers of larger spiral galaxies but not from progenitors
of faint ellipticals.  Because the feedback strength depends on the
mass of galaxies, the metallicity of stars in more massive progenitors
becomes higher than that in small progenitors.  Therefore massive
ellipticals become redder than small ellipticals.  However, in the
hierarchical clustering scenario, larger objects are formed by mergers
of smaller objects.  So such massive progenitors are formed by mergers
of smaller galaxies.  Thus the above conclusion may be insufficient
for accounting to make the CMR, and it is needed to investigate how
the metal abundance evolves in galaxies.  The aim of this paper is,
therefore, exploring what process, such as feedback and star
formation, affects the CMR and the evolution of metallicity.
Especially we focus on the slope of the CMR.

In Section 2, we describe the semi-analytic model used here briefly.
In Section 3, we show the colour--magnitude relation in almost the same
model as that of KC, and show evolutions of mean stellar metallicities
of galaxies.  In Section 4, we investigate the parameter dependence
and some effects on the CMR.  Section 5 is devoted to conclusions and
discussion.

\section{MODEL}
In this section, the semi-analytic model which we use is described.
Since the semi-analytic model includes some physical processes
connected with the galaxy formation, it is easy to understand the
galaxy formation process, in contrast to N-body and hydrodynamical
simulations.  Therefore we must include the following ingredients in
our models at least: merging history of dark haloes, cooling and
heating processes of baryonic gas, star formation and feedback by
supernovae, mergers of galaxies, and stellar population synthesis
model.  The following procedure is mainly based on Kauffmann, White \&
Guiderdoni (1993), Cole et al. (1994) and Somerville \& Primack
(1998).

The outline of the procedure is as follows.  At first, the merging
paths of dark haloes are realized by the extension of the PS
formalism.  Next, in each merging path, evolution of the baryonic
component, namely, gas cooling, star formation, and supernova
feedback, are calculated.  We recognize a system consisting of the
stars and cooled gas as a {\it galaxy}.  When two or more dark haloes
merge together, there is a possibility that galaxies contained in
progenitor haloes merge together.  The judgment whether the galaxies
merge together or not is based on the dynamical friction time-scale.
Here we define the {\it lifetime} of haloes as the elapsed time
between the formation time of the halo and the time at which the halo
is subsumed into a larger halo.  If the dynamical friction time-scale
is shorter than the lifetime, the galaxies will merge.  If not, the
galaxies will not merge and the common dark halo has two or more
galaxies.  Finally, we calculate the colour and luminosity of each
galaxy from the star formation history of each galaxy.  Through the
above procedures, we obtain a colour--magnitude diagram of galaxies.

\subsection{Merging history of dark haloes}
The merging histories of dark haloes are realized by using the method
developed by Somerville \& Kolatt (1999).  They improved the model by
Kauffmann \& White (1993), based on the extension of the
Press--Schechter formalism (Press \& Schechter 1974) by Bower (1991),
Bond et al. (1991) and Lacey \& Cole (1993), for satisfying the mass
conservation of haloes.

Their algorithm is simple and can be quickly performed as follows.

First, we consider two density fluctuation fields with two different
smoothing scales, which corresponds to mass scales $M_{1}$ and $M_{0}$
($M_{1}<M_{0}$).  The variances of the density fluctuation fields are
$S_{1}$ and $S_{0}$, respectively.  The variance $S_{i}$ corresponds
to mass scale $M_{i}$ uniquely if we fix a power spectrum of the
density fluctuation field.  The linear density contrast collapsing
just at $z_{i}$ is obtained as $\omega_{i}=1.69(1+z_{i})$ by the
spherically symmetric collapse approximation (Tomita 1969; Gunn \&
Gott 1972).

Next we consider the probability distribution function of $\Delta
S=S_{1}-S_{0}$, $p(\Delta S,\Delta\omega)d\Delta S$, where
$\Delta\omega=\omega_{1}-\omega_{0}$.  When the density fluctuation
fields are Gaussian random fields, this probability is simply
\begin{equation}
p(\Delta S,\Delta\omega)d\Delta
S=\frac{1}{\sqrt{2\pi}}\frac{\Delta\omega}{(\Delta S)^{3/2}}
\exp\left[-\frac{(\Delta\omega)^{2}}{2\Delta S}\right]d\Delta S.
\label{eqn:prob}
\end{equation}
This function is a simple Gaussian distribution, if we change the
variable to $x\equiv\Delta\omega/\sqrt{\Delta S}$.  After this, we
interpret the variance $S$ as the mass $M$.

The following procedure is based on a Monte Carlo method.  First we
set a mass $M_{0}$, or $S_{0}$, as a mass of a finally collapsing
halo, e.g., a cluster scale or the `Milky Way' scale, and `time-step'
$\Delta\omega$.  By using eq.(\ref{eqn:prob}), we pick out $\Delta S$
randomly.  From this $\Delta S$, $S_{1}$, that is, $M_{1}$, is
determined.  This $M_{1}$ is interpreted as a mass of a progenitor
halo of the halo with mass $M_{0}$, collapsing at $z=z_{1}$ estimated
by $\Delta\omega$.  If $M_{1}\leq M_{l}$, where $M_{l}$ is a minimum
mass for an object identified as an isolated halo at a corresponding
redshift, the mass $M_{1}$ is added to a diffuse `accretion mass'.

The above process is repeated until the rest of mass of the halo
becomes less than $M_{l}$.  In the same way, merging paths are
recursively realized by regarding the halo with mass $M_{1}$ as a
starting point halo such as the halo with mass $M_{0}$.

Mass functions at various redshifts obtained by using the above
procedure reproduce the Press--Schechter mass function well.  Of
course, the Press--Schechter formalism itself has some problems, e.g.,
this formula does not include the spatial correlation of density
fluctuations (Yano, Nagashima \& Gouda 1996; Nagashima \& Gouda 1997).
Porciani et al. (1998) consider the effect of the spatial correlation
in their method for realizing merging paths approximately.  However,
in spite of many theoretical uncertainties, the difference between the
Press--Schechter mass function and the mass functions given by N-body
simulations is only a few factor in the number of haloes.  This
curiosity is reviewed by Monaco (1998).  Therefore, for simplicity, we
use the extension of the PS formalism in this paper.

In this paper, we realize merging paths of progenitors of haloes with
mass $5\times 10^{14}$M$_{\odot}$, which corresponds to the circular
velocity $V_{c}=10^{3}$ km~s$^{-1}$.  This is a typical size of
clusters of galaxies.  The time-steps are fixed to $\Delta z=0.05$.
The minimum mass for identifying as an isolated halo is
$M_{l}=10^{10}$M$_{\odot}$.  Smaller objects than $M_{l}$ may not be
able to collapse because the mass of such objects are smaller than the
Jeans mass ($\sim 10^{10}$M$_{\odot}$) after the reionization of
intergalactic medium by the UV background radiation at $z\sim 10$
(Ostriker \& Gnedin 1996), and because galaxies are formed actively at
$z\la 5$.  We find that results presented in this paper does not
depend on the choice of $M_{l}$ $\la 10^{10}$M$_{\odot}$.  The power
spectrum $P(k)$ of density fluctuations given by Bardeen et al. (1986)
is adopted,
\begin{eqnarray}
\lefteqn{P(k)\propto kT^{2}(k),}\\
\lefteqn{T(k)=\frac{\ln(1+2.34q)}{2.34q}\times}\nonumber\\
&&\left[1+3.89q+(16.1q)^{2}
+(5.46q)^{3}+(6.71q)^{4}\right]^{-1/4},\\
\lefteqn{\quad ~q\equiv\frac{k}{\Omega h^{2} \mbox{Mpc}^{-1}},}
\end{eqnarray}
where $\Omega$ is the cosmic density parameter and $h$ is the Hubble
parameter, $H_{0}=100h$ km~s$^{-1}$~Mpc$^{-1}$.  We adopt the standard
CDM model, $\Omega=1$ and $h=0.5$, in this paper.

\subsection{Gas cooling}\label{sec:cool}
In each realized merging path, we evaluate the amount of cooled gas
through the radiative cooling, that is a direct material for forming
stars.

When a dark halo under consideration collapses, we estimate the mass
of hot diffuse gas contained in the halo.  The sources of the diffuse
gas are hot gas contained in its progenitor haloes and in the
accreting matter.  The baryon fraction of the accreting matter is
defined as $\Omega_{b}/\Omega$, where $\Omega_{b}$ is the cosmic mean
baryon density.  The density distribution $\rho(r)$ of this hot gas in
the halo is assumed as the isothermal distribution, $\rho(r)\propto
r^{-2}$.  The absolute value of the density profile is determined by
mass and collapsing redshift of haloes under the spherical collapse
model.

The cooling time-scale $\tau_{cool}$ is obtained as a function of the
radius from the hot gas density profile, the temperature of the hot
gas, and the cooling function $\Lambda(T)$ as follows,
\begin{equation}
\tau_{cool}(r)=\frac{3}{2}\frac{\rho(r)}{\mu
m_{p}}\frac{kT}{n_{e}^{2}(r)\Lambda(T)},
\label{eqn:cltime}
\end{equation}
where $\mu m_{p}$ is the mean molecular weight, $n_{e}(r)$ is the
electron number density at the radius $r$, and $k$ is the Boltzmann
constant.  The cooling function depends on the metallicity of the hot
gas.  We adopt the both cooling functions of zero-metallicity and
solar-metallicity given by Sutherland \& Dopita (1991), and determine
the cooling efficiency depending on the metallicity of the gas from
the two cooling functions by interpolation and extrapolation.

When a halo merges with another halo of comparable size (when the mass
ratio of these haloes is larger than $f_{reheat}$), the diffuse hot
gas contained in the halo is shock heated to the virial temperature of
the halo.  In this case, the hot gas within the {\it cooling} radius
$r_{cool}$ cools and the cooled gas mass is simply added to the cold
gas reservoir of the galaxy.  The cooling radius is a radius at which
the cooling time-scale equals to the time-step $\Delta t$ of the
merging histories of dark haloes.  On the other hand, when the halo
merges with small haloes whose mass ratio to the halo is less than
$f_{reheat}$, the hot gas is not shock heated and preserves its
temperature, and the density profile is not disturbed.  This parameter
is introduced by Somerville \& Primack (1998) firstly.  In such a
case, gas contained in the smaller haloes may remain in outer region
of the new halo.  Therefore, we introduce the following process when
the mass ratio of the merging haloes is less than $f_{reheat}$.  Now
we define $t_{elapse}$ as the elapsed time since the last shock heated
epoch of the halo and $r_{cool}(t_{elapse})$ as a radius at which the
cooling time-scale equals to the elapsed time.  In this case, because
the density profile is maintained and the gas within
$r_{cool}(t_{elapse})$ has already cooled, the cooled gas mass is
estimated as
\begin{equation}
\Delta
M_{cool}=\int_{r_{cool}(t_{elapse})}^{r_{cool}(t_{elapse}+\Delta
t)}\rho(r)4\pi r^{2}dr.
\label{eqn:clint}
\end{equation}
This procedure is adopted only when $r_{cool}(t_{elapse})/R\leq 0.8$.
When $r_{cool}/R>0.8$, the hot gas may fall into the centre of haloes
and cool because most of hot gas has already cooled at the previous
time-step.  So the gas density profile is recalculated and $r_{cool}$
is estimated by $\tau_{cool}=t_{life}$ as in the case of the major
merger of haloes, while the density profile of the dark halo will be
preserved.

We can avoid the dependence of the amount of the cooled gas mass on
the artificial time-step of calculation by using the above procedure.
In this paper, we use $f_{reheat}=0.2$.  Of course, if we use the
block model which was used by the Durham group (Cole \& Kaiser 1988;
Cole et al. 1994), this parameter is not required.  In the extension
of the PS formalism, masses of haloes are given by randomly and the
time-step is fixed.  On the other hand, in the block model, density
contrasts which determine the time-step are given randomly, while
masses of haloes are divided by a factor of 2.  So mass of a new
collapsing halo is always twice of a maximum progenitor.  Thus the
mass ratio of the merging haloes are usually larger than
$f_{reheat}=0.2$.

In the case that the circular velocity of haloes exceeds 500 km
s$^{-1}$, we prevent the cooling process by hand to suppress the
formation of extremely big {\it monster} galaxies.  If we consider
only the mean density of the hot gas, the maximum mass of galaxies is
$\sim 10^{12}$M$_{\odot}$ (Rees \& Ostriker 1977).  In our
calculation, since we adopt the isothermal distribution
$\rho(r)\propto r^{-2}$, the cooling process becomes efficient in the
central region of haloes.  Actually this phenomenon is suggested
observationally (e.g., Fabian 1994).  However the number density of
bright galaxies is much larger than the observed one, unless we
prevent the cooling process in such large haloes.  Thus we adopt the
same unsatisfactory solution as Kauffmann et al. (1993).  Somerville
\& Primack (1998) also adopt the same solution but for $400$
km~s$^{-1}$ of the cut-off circular velocity.  A detailed study is
needed in order to understand this effect physically.

\subsection{Star formation}\label{sec:sf}
In Kauffmann et al. (1993), the star formation in disc is described by
the following simple law,
\begin{eqnarray}
\dot{M_{*}}&=&\frac{M_{cold}}{\tau_{*}},\label{eqn:sfr}\\
\tau_{*}&=&\tau_{*}^{0}(1+z)^{-3/2},
\label{eqn:sf}
\end{eqnarray}
where eq.(\ref{eqn:sfr}) denotes the rate of stars newly formed.  This
form is the same as the one adopted by Kauffmann et al. (1993), and
referred as the `Munich model' by Somerville \& Primack (1998).  The
redshift dependence on the star formation time-scale $\tau_{*}$ is
obtained as follows.  We adopt the simple star formation
interpretation, $\tau_{*}\propto R/V_{c}$, that is, $\tau_{*}$ is
proportional to the dynamical time-scale.  From this, the virial
theorem and the spherical collapse model, we obtain that
$\tau_{*}\propto (1+z)^{-3/2}$.  The value of the star formation
time-scale at $z=0$, $\tau_{*}^{0}$, is a free parameter.  This
parameter describes the star formation with a long time-scale in a disc
of spiral galaxies.  Moreover we need to fit the amount of the cold
gas in the `Milky Way'-like spiral galaxies, which have the same
circular velocity as our Galaxy (Milky Way), $V_{c}\simeq 220$
km~s$^{-1}$, to the amount of the cold gas in our Galaxy.  If
$\tau_{*}^{0}$ is small, almost all of the cold gas is transformed
into the stars until the present epoch.  So $\tau_{*}^{0}$ is larger
than a few Gyr.

Stars formed by the above process constitute a disc of galaxies.  Star
formation in a bulge occurs when two or more galaxies with comparable
masses merge together (see Section \ref{sec:merger}).  In this merging
process, since the system is dynamically disturbed, starburst may
occur.  Then both stars formed in the burst and disc stars may fall
into the central region of the galaxy and constitute the bulge.  We
describe this event by adopting very short star formation time-scale.

Of course, the above star formation laws are oversimplified.  In the
present situation, stars are considered to be formed as follows.  The
star formation process may be determined by the local properties, such
as density and temperature.  If we consider only atomic cooling in our
calculation, the gas cools until $\sim 10^{4}$ K.  So we identify such
gas cooled until $\sim 10^{4}$ K as the cold gas.  However, the
pressure of the gas is still too high to form stars.  The cooling by
molecules is needed to cool the gas lower than $10^{4}$ K.  Then the
gas cools until a few K by the molecular cooling, and fragments into
small clumps.  These clumps evolve to stars.  These processes are
performed in molecular clouds, whose mass is about
$10^{6}$M$_{\odot}$.  Therefore we should consider the processes of
the molecular cooling and the fragmentation of the `cold' gas ($\sim
10^{4}$ K).  It is studied by considering the ensemble of molecular
clouds to connect the star formation in molecular clouds with the
global star formation properties in galaxies (Fujita 1998; Fujita \&
Nagashima 1999).  These processes are very complicated, so we do not
adopt them in our semi-analytic model, since the simplicity which is
an important merit of this model will be lost.

\subsection{Supernova feedback}

When stars form, massive stars with short lifetime explode and heat up
the surrounding cold gas.  The number of supernovae per solar mass of
stars formed is $\eta_{SN}\simeq 7\times 10^{-3}$M$_{\odot}^{-1}$ for
the Salpeter IMF (Salpeter 1955).  The kinetic energy from a
supernova, $E_{SN}$, is about $10^{51}$ erg.  When a fraction
$\epsilon$ of the released energy is used to heat up surrounding cold
gas, the amount of this reheated gas is evaluated as
\begin{equation}
\Delta M_{reheat}=\epsilon
\frac{4}{5}\frac{\dot{M_{*}}\eta_{SN}E_{SN}}{V_{c}^{2}}\Delta t.
\label{eqn:sn}
\end{equation}
This feedback process has many uncertainties actually, and therefore
we adopt a simple description (Cole et al. 1994),
\begin{equation}
\Delta M_{reheat}=\left(\frac{V_{c}}{V_{hot}}\right)^{-\alpha_{hot}}
\dot{M_{*}}\Delta t\equiv \beta \dot{M_{*}}\Delta t,
\label{eqn:alphot}
\end{equation}
where $V_{hot}$ and $\alpha_{hot}$ are free parameters.
Eq.(\ref{eqn:sn}), which is adopted by Kauffman et al. (1993)
corresponds to $\alpha_{hot}=2$, and Cole et al. (1994) used
$\alpha_{hot}=5.5$ in their fiducial model.

It is the conclusion of KC that this feedback process is a key
parameter to determine the slope of the CMR.  Therefore we also
investigate the dependence of CMR on the feedback process.

\subsection{Mergers of galaxies}\label{sec:merger}
When two or more haloes merge together, the new common halo may have
two or more galaxies.  In that case, the galaxies will lose their
energy due to dynamical friction, and then fall into a centre of the
new common halo.  Finally they may merge together as the case may be.

When a halo collapses, a central galaxy contained in the largest
progenitor is identified as the central galaxy of the new common halo.
The cooled gas described in Section \ref{sec:cool} accretes to this
central galaxy.  Other galaxies are identified as satellite galaxies.

We calculate the elapsed time from the epoch at which each satellite
galaxy was identified as {\it satellite}, namely, it was contained in
a common halo not as a central galaxy but as a satellite galaxy.  It
should be noted that this elapsed time is reset to 0 when the `major
merger' of dark haloes occurs (see Section \ref{sec:cool}) because
orbits of satellite galaxies may be violently disturbed by the major
merger of haloes.  When this elapsed time exceeds the dynamical
friction time-scale for the galaxies in the new common halo, the
satellite merges with the central galaxy.  The dynamical friction
time-scale is (Binney \& Tremaine 1987)
\begin{equation}
\tau_{mrg}=\frac{1.17 R^{2}V_{c}}{\ln\Lambda GM_{sat}},
\label{eqn:dynf}
\end{equation}
where $R$ and $V_{c}$ are the virial radius and circular velocity of
the new common halo, respectively, $M_{sat}$ is the total mass of a
halo to which the satellite belonged as central galaxy, and
$\ln\Lambda$ is the Coulomb logarithm, which is approximated as
$\simeq\ln (1+M_{H}^{2}/M_{sat}^{2})$ ($M_{H}$ is the mass of the
common halo) (Somerville \& Primack 1998).  

In this paper, we do not treat mergers between satellite galaxies for
simplicity, which is described in Somerville \& Primack (1998).  There
is a possibility that this effect changes the resulting CMRs.  This
effect will be considered in future work.

When a satellite galaxy merges with a central galaxy, and moreover the
mass ratio of the galaxies is larger than $f_{bulge}$, all stars of
the satellite galaxy and disc stars of the central galaxy are
incorporated with the bulge of the central galaxy.  Then cold gases of
both galaxies turn to stars in the bulge with a very short time-scale
(`starburst').  In this paper, we use the following two burst model
on the feedback process.  One is that the strength of the feedback is
the same as that in the disc, according to eq.(\ref{eqn:alphot})
(model sbA).  Another is that there is no feedback effect during the
burst, that is, all cold gas turns to stars through the `closed-box'
process (model sbB).  In order to decide which model is correct,
detailed simulations of merging process will be needed.  On the other
hand, if the mass ratio of the galaxies is smaller than $f_{bulge}$,
all stars of the satellite galaxy are incorporated with the disc of
the central galaxy and the cold gas reservoir of the satellite merges
with that of the central galaxy, and then disc stars are formed from
the cold gas with the star formation time-scale obtained from
eq.(\ref{eqn:sf}).  In this paper, we adopt $f_{bulge}=0.2$.  This
value is the same as that in Somerville \& Primack (1998).  They adopt
this value for reproducing observational fractions of the number of
galaxies of each morphology.

\subsection{Chemical evolution}
\begin{figure}
\epsfxsize=8cm
\epsfbox{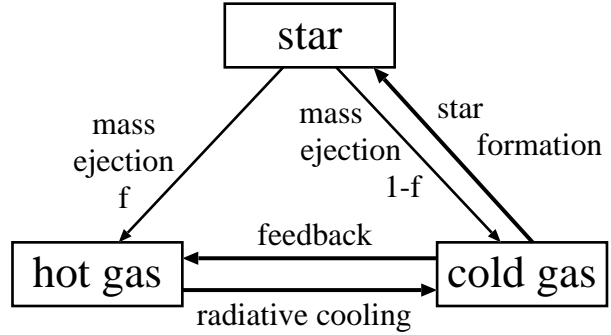}
\caption{Schematic description of gas and metal recycling in star
  formation.  $f$ is a fraction of ejected mass from evolved stars
  into hot gas directly.}
\label{fig:chem}
\end{figure}

Chemical evolution is treated in almost the same way as described in
KC.  The instantaneous recycling approximation is adopted (Tinsley
1980).  We show the gas and metal exchange in Fig.\ref{fig:chem}
schematically.  Metals ejected from supernovae are recycled as
follows.

When two or more haloes merge together, metals contained in the hot
gas is joined together.  If two or more galaxies merge together,
metals contained in the cold gas reservoirs of galaxies are joined
together.  When the hot gas cools, metals contained in the cooled gas
are also joined to the cold component.  Here we assume that metals in
cold and hot components are well mixed in each component.  Stars
formed from the cold gas have initially the same metallicity as the
cold gas.  This metallicity affects colours of galaxies directly.

The amount of metals ejected from supernovae is characterized by $y$,
which is heavy element yield for each generation of stars.  The
fraction $f$ of the ejecta is ejected directly into the hot gas, and
the rest, $1-f$, is incorporated with the cold gas.  We adopt $f=0.3$,
which is the same value as in KC.  The gas fraction returned by
evolved stars $R$ is 0.75 in this paper.  Simultaneously, the
supernovae heat up the surrounding cold gas, therefore metals
contained in the cold gas are also returned to the hot gas.

\subsection{Stellar population synthesis and identification of morphologies}
\label{sec:mor}

In order to compare our results with observations such as CMRs
directly, stellar population synthesis model must be considered.  We
adopt a model by Kodama \& Arimoto (1997) as such a population
synthesis model.  Luminosity and colour of stars cannot be
theoretically estimated without uncertainties because the stellar
evolution models themselves include the uncertainties.  For example,
as for the convection of stellar gas, we have only a phenomenological
theory (`mixing length theory').  However, their model is adequate for
our purpose of investigating the slope of the CMR.  Once we understand
the properties of galaxies such as the CMR physically and
qualitatively, even if the stellar population model is changed, we fit
immediately our results with observations by changing parameters
mentioned above.

The IMF which we adopt is the Salpeter type with a slope of 1.35, and
the mass range is $0.1$M$_{\odot}\sim 60$M$_{\odot}$.  The range of
stellar metallicity $Z_{*}$ of simple stellar populations is
$0.0001\sim 0.05$.

In Section \ref{sec:sf}, we divide the stellar component into disc and
bulge components.  Morphology of each galaxy is determined by the
$B$-band bulge-to-disc luminosity ratio ($B/D$).  Simien \& de
Vaucouleurs (1986) showed that the Hubble type of galaxies correlates
with the $B$-band luminosity $B/D$.  In this paper, galaxies with
$B/D\geq 1.52$ are identified as ellipticals, $0.68\leq B/D<1.52$ as
S0s, and $B/D<0.68$ as spirals, according to their results.  It is
shown that this method for classification reproduces observations well
by Kauffmann et al. (1993) and Baugh, Cole \& Frenk (1996).

\section{Origin of colour--magnitude relation}\label{sec:result}
In this section, we explore the origin of the CMR.  In this paper,
because we investigate how the CMR depends on the physical processes
such as star formation, supernova feedback, and so on, we fix the
cosmological parameters to the standard CDM model, that is, $\Omega=1,
\Lambda=0, H=50$km s$^{-1}$ Mpc$^{-1}$, $\Omega_{b}=0.06$, and $
\sigma_{8}=0.67$.  We refer to the models considered by the models A,
B, C and D as shown in Table 1.  In this table, we also show models
from E to J, which are adopted in the next section.  The fifth column,
UV, is explained in Section \ref{sec:uv}.  The sixth column, burst, is
shown in Section \ref{sec:merger}.  The yield $y$ is equal to
$0.038=2Z_{\odot}$ in all the models.  In KC, $y=1.2Z_{\odot}$ in the
low feedback model.  However, in order to see the effects of the
feedback and so on, we fix the value of the yield in all models.

In the following sections, we take notice of only the slopes of CMRs,
because the luminosity of galaxies can be translated by considering
the following reason.  Stars are formed according to the IMF.  Mass of
luminous stars is larger than $\sim 0.08$M$_{\odot}$, which is
determined by the criterion of nuclear burning.  However, there is a
possibility that invisible stars with mass smaller than
$0.08$M$_{\odot}$ are formed.  If there are many invisible stars in
galaxies, the galaxies become faint compared to the case that all
stars are luminous.  The ratio of the invisible stars to the luminous
stars is treated as a free parameter in the previous work.  Therefore
the absolute value of luminosity can be adjusted by this parameter.
The suitable value of this parameter will be determined by considering
other observational quantities.  Thus this parameter does not affect
the slope of the CMR.

\begin{table}\label{tab}
\begin{center}  
\caption{Parameters for models.}
\label{tab:param}
\begin{tabular}{cccccc}
Model & $V_{hot}$ (km~s$^{-1}$) & $\alpha_{hot}$ & $\tau_{*}^{0}$
(Gyr) & UV & burst\\
A & 280 & 2 & 20 & off & sbA\\
B & 100 & 2 & 20 & off & sbA\\
C & 280 & 2 & 20 & off & sbB\\
D & 100 & 2 & 20 & off & sbB\\
E & 280 & 2 & 2 & off & sbB\\
F & 100 & 2 & 2 & off & sbB\\
G & 240 & 5.5 & 2 & off & sbB\\
H & 100 & 5.5 & 2 & off & sbB\\
I & 400 & 2 & 2 & off & sbB\\
J & 280 & 2 & 2 & on & sbB
\end{tabular}
\end{center}
\end{table}

\subsection{Colour--magnitude relations}

\begin{figure}
\epsfxsize=8cm
\epsfbox{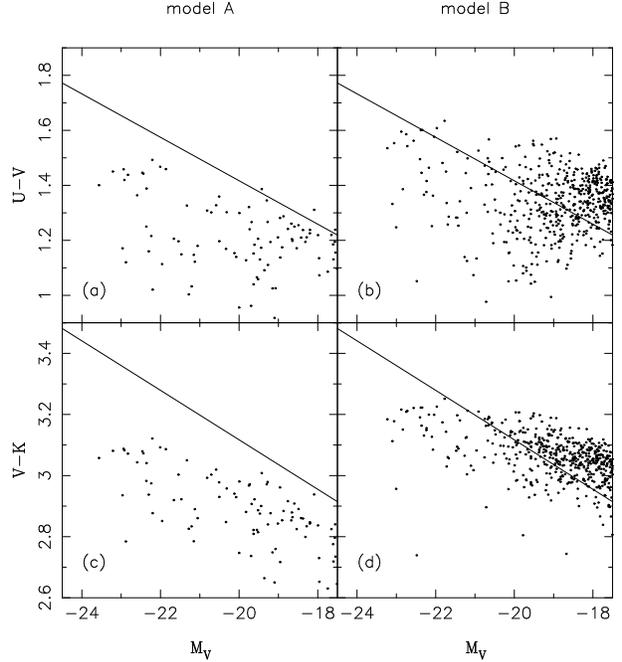}
\caption{Colour--magnitude diagram of model A and B.  (a) $M_{V}$
  v.s. $U-V$ in the model A.  (b) Same as (a), but for the model B.
  (c) $M_{V}$ v.s. $U-V$ in the model A.  (d) Same as (c), but for the
  model B.  The models are characterized in Table 1.  The dots denote
  galaxies identified as ellipticals.  The solid lines show the
  observed CMRs (Bower et al. 1992).}
\label{fig:cmrab}
\end{figure}

In Fig.\ref{fig:cmrab}, we show the CMRs in the models A and B with
$\alpha_{hot}=2$ which is the value adopted by KC.  The dots denote
galaxies identified as ellipticals, and the solid lines show the
observational CMRs (Bower et al. 1992).  The criterion to pick out
ellipticals among all galaxies is shown in Section \ref{sec:mor}.

As in KC, in the model A with high feedback efficiency, the slopes of
the CMRs are in roughly agreement with the observations, but the
dispersion is larger than that of observations, $\sim 0.04$ mag.  This
is due to the recent star formation owing to long star formation
time-scale, $\tau_{*}^{0}=20$ Gyr.  On the other hand, in the model B
with low feedback efficiency, the slopes of the CMRs are nearly flat.
At the bright-end, the colour becomes redder about 0.1-0.2 mag, and at
the faint-end ($M_{V}\sim -18$), about 0.2-0.3 mag, compared to that
in the model A.  In order to see the physical relation between the
reddening and the feedback intensity, we investigate the age- and
metallicity-luminosity relations.

\begin{figure}
\epsfxsize=8cm
\epsfbox{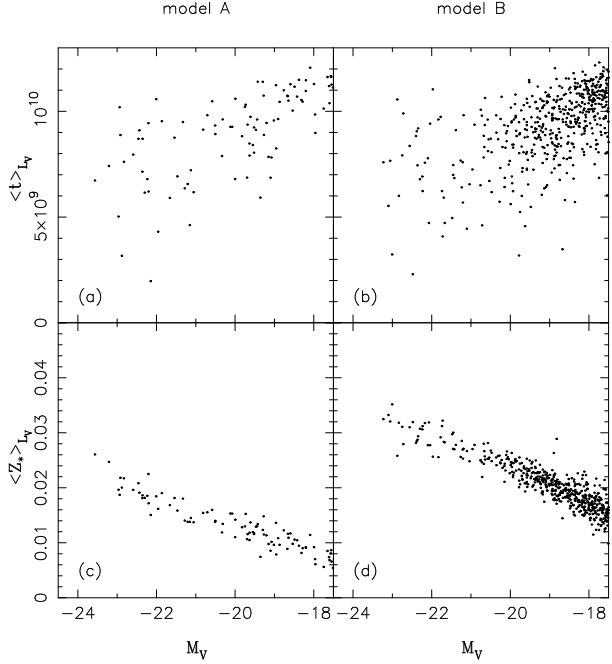}
\caption{Upper panels: age-magnitude diagram.  (a) Model A.  (b) Model
  B.  Lower panels: metallicity-magnitude diagram.  (c) Model A.  (d)
  Model B.  The dots denote elliptical galaxies.}
\label{fig:mtab}
\end{figure}

We define the V-luminosity weighted mean stellar age and
metallicity as follows,
\begin{eqnarray}
\langle Z_{*}\rangle_{L_{V}}&\equiv&\frac{
\int_{0}^{t_{0}} L_{V}(t,Z_{cold})\dot{M_{*}}(t)Z_{cold}(t)dt}
{\int_{0}^{t_{0}} L_{V}(t,Z_{cold})\dot{M_{*}}(t)dt},\\
\langle t\rangle_{L_{V}}&\equiv&\frac{
\int_{0}^{t_{0}} L_{V}(t,Z_{cold})\dot{M_{*}}(t)tdt}
{\int_{0}^{t_{0}} L_{V}(t,Z_{cold})\dot{M_{*}}(t)dt},
\end{eqnarray}
where $Z_{cold}$ is the metallicity of cold gas, $\dot{M_{*}}(t)$ is
the star formation rate at $t$, and $L_{V}(t,Z_{cold})$ is the
luminosity of stars at present, $t=t_{0}$, with age $t_{0}-t$ and
metallicity $Z_{cold}$.  In both models, the distributions of mean
stellar age are alike (upper panels of Fig.\ref{fig:mtab}), and the
mean stellar ages spread about $5\sim 11$ Gyr.  On the other hand,
there are tight relations between the V-luminosity and the mean
stellar metallicity.  The slopes of the relations in the models A and
B are almost the same, but the value of the metallicity in the model B
is higher than that in the model A.  In the model B, because ages of
most of dwarf ellipticals are old and because the metallicity is high,
the CMRs are flatter than those in the model A.

\begin{figure}
\epsfxsize=8cm
\epsfbox{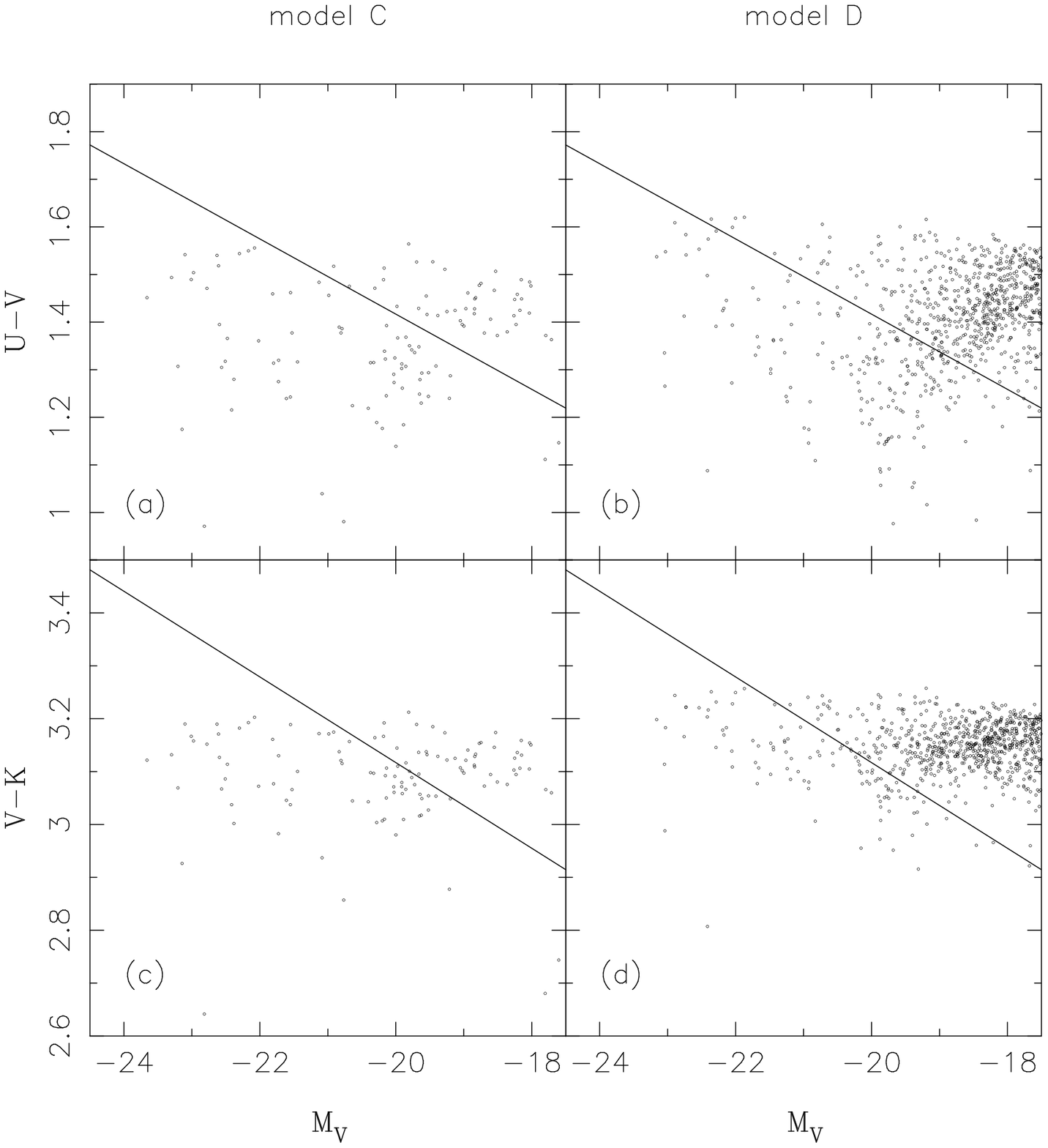}
\caption{Same as Fig.\ref{fig:cmrab}, but for model C and D.
  Both models have a long star formation time-scale, $\tau_{*}^{0}=20$
  Gyr, and the starburst model sbB is adopted.  Model C: $V_{hot}=280$
  km~s$^{-1}$.  Model D: $V_{hot}=100$ km~s$^{-1}$.}
\label{fig:cmrcd}
\end{figure}

\begin{figure}
\epsfxsize=8cm
\epsfbox{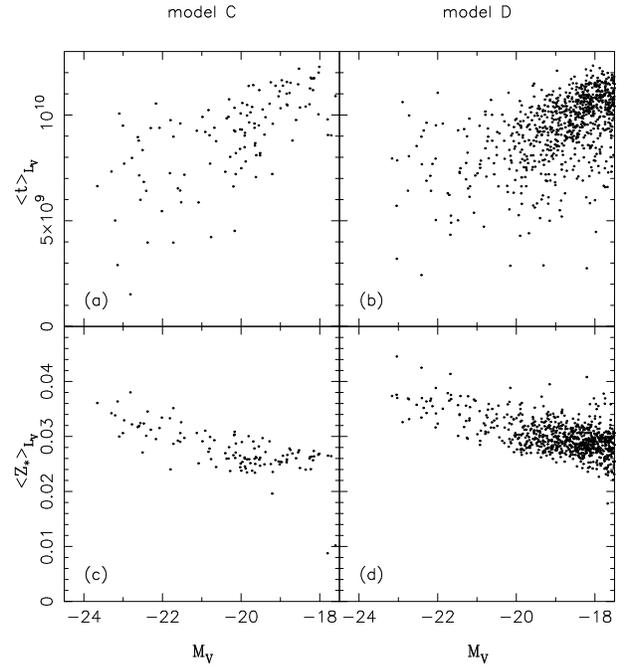}
\caption{Same as Fig.\ref{fig:mtab}, but for model C and D.}
\label{fig:mtcd}
\end{figure}

Next we show the CMRs in the models C and D, in which the all cold gas
is transformed to stars in the starburst process
(Fig.\ref{fig:cmrcd}).  In both models, the slopes of CMRs are flat,
as well as in the model B, while the number of galaxies in the model C
is less than that in the model D, owing to the high feedback
efficiency in the model C.  The difference between the models A and C
is only in the feedback model.  This difference will cause the
difference in the metallicity of galaxies (see Section
\ref{sec:feedback}).  In Fig.\ref{fig:mtcd}, We show the same figure
as Fig.\ref{fig:mtab} but for the model C and D.  The age
distributions of the models C and D are almost the same as those of
the models A and B.  On the other hand, the metallicity distributions
are different.  The mean metallicities of dwarf ellipticals increase
by $\langle Z_{*}\rangle_{L_{V}}\sim 0.03$ and the distributions are
flat in both models C and D, while $\langle Z_{*}\rangle_{L_{V}}\sim
0.01$ at $M_{V}\sim-19$ in the model A.  Therefore the difference of
the feedback models mainly affects the metallicity of dwarf
ellipticals.  Thus we can say that the feedback model is essential to
determine the properties of the CMR, when $\tau_{*}^{0}=20$ Gyr (see
Section \ref{sec:feedback}).

From these figures, we conclude that the CMR results from the
metallicity-luminosity relation as KC concluded.  The conclusion of
the origin of the metallicity-luminosity relation in KC is that
ellipticals are formed by mergers of gas-rich spirals, and that
massive ellipticals are formed from massive spirals because the
feedback efficiency of such massive galaxies is small, and so more
metal-rich stars are formed (Fig.2 in KC).  We will show the
evolutions of mean stellar metallicity and age of each galaxy in order
to investigate this process in detail.

\subsection{Evolution of mean stellar metallicity}

In this subsection, we show the evolutions of the mean stellar
metallicity $\langle Z_{*}\rangle_{M_{*}}$, which are weighted by
stellar mass for simplicity.

\begin{figure}
\epsfxsize=8cm
\epsfbox{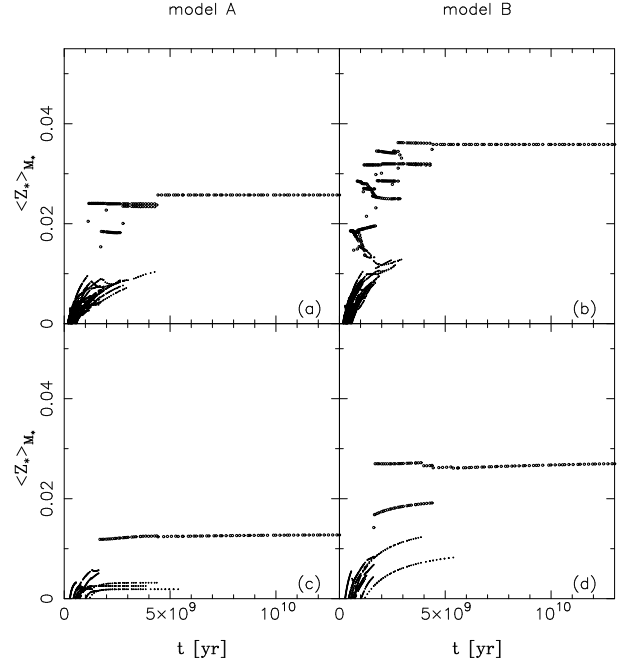}
\caption{Evolution of mass-weighted mean stellar metallicity.  Each
symbol shows the mean metallicity of a progenitor at the time.  The
open circles denote progenitors in the {\it elliptical state}, and the
dots the {\it spiral state}.  These states are defined in the text.
(a) A luminous elliptical in model A.  $M_{V}=-22.0, U-V=1.47,
V-K=3.09$.  (b) A luminous elliptical in model B.  $M_{V}=-22.0,
U-V=1.61, V-K=3.23$.  (c) A faint elliptical in model A.
$M_{V}=-18.4, U-V=1.24, V-K=2.86$.  (d) A faint elliptical in model B.
$M_{V}=-19.1, U-V=1.31, V-K=3.07$.}
\label{fig:idab}
\end{figure}

In the upper panels of Fig.\ref{fig:idab}, we show the evolutions of
stellar metallicity of luminous ellipticals picked out in the models A
and B.  Their final magnitudes and colours are $M_{V}=-22.0, U-V=1.47,
V-K=3.09$ and $M_{V}=-22.0, U-V=1.61, V-K=3.23$, respectively.  The
open circles denote the mean stellar metallicities of progenitors in
the {\it elliptical state}, which is defined as the bulge-to-disc
stellar mass ratio $M_{bulge}/M_{disc}\geq 0.6$, and the dots show
those in the {\it spiral state}, $M_{bulge}/M_{disc}<0.6$.  These two
luminous galaxies are identified as galaxies experiencing the same
merging path of dark haloes.  Because the merging histories of dark
haloes of these galaxies are the same, the difference between results
in different models is caused from only parameters shown in Table
\ref{tab:param}.  These galaxies rapidly grow in mass and the feedback
effect becomes weak at early times.  So the mean stellar metallicities
also exceeds $Z_{\odot}$ at early times.  Therefore, the difference
between these galaxies is small.  The final metallicity of elliptical
galaxy in the model A (Fig.\ref{fig:idab}a) is smaller than the yield
$y=0.038$.  The reason for this is considered as follows.  One is that
the feedback effect remains a little because of the shallower
dependence of feedback on circular velocity, $\alpha_{hot}=2$.  Second
is that the star formation time-scale is very long, $\tau_{*}^{0}=20$
Gyr.  On the other hand, the difference between faint ellipticals is
large (lower panels of Fig.\ref{fig:idab}).  The ratio of the final
metallicities is more than a factor of two.  These faint ellipticals
are also picked out in the same merging path.  Their final magnitudes
and colours are $M_{V}=-18.4, U-V=1.24, V-K=2.86$ and $M_{V}=-19.1,
U-V=1.31, V-K=3.07$, respectively.  In the high feedback efficiency
model A, the growth of the mean stellar metallicity is slow and the
final metallicity is very small, $\langle Z_{*}\rangle_{M_{*}}\sim
0.013$.  On the other hand, the growth in the low feedback efficiency
model B is rapid and then the final metallicity is high, $\langle
Z_{*}\rangle_{M_{*}}\sim 0.028$.  This shows that in the low feedback
model even small galaxies can evolve chemically.

In this figure, some progenitors have periods of decrease of their
metallicities.  This is because gas with low metallicity accretes a
galaxy with high mean metallicity and stars are newly formed from the
gas.  Therefore such stars have low metallicity and the mean
metallicity of the galaxy decreases.

\begin{figure}
\epsfxsize=8cm
\epsfbox{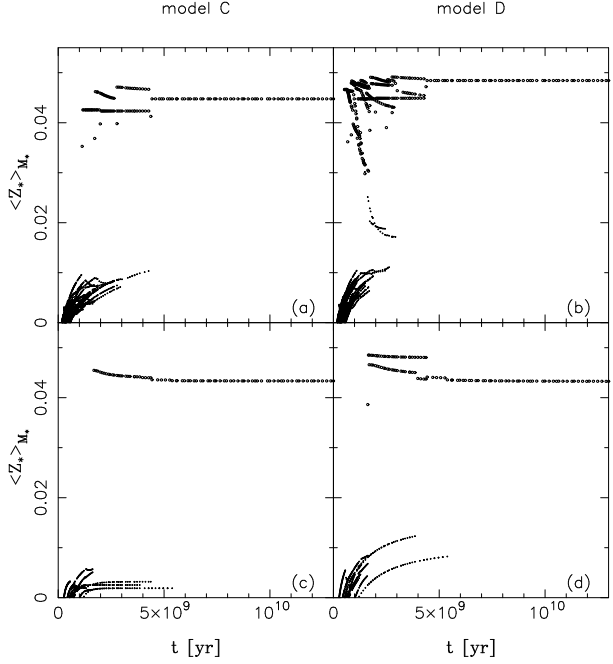}
\caption{Same as Fig.\ref{fig:idab}, but for model C and D.  (a)
  $M_{V}=-22.2, U-V=1.55, V-K=3.19$ in model C.  (b) $M_{V}=-22.0,
  U-V=1.62, V-K=3.25$ in model D.  (c) $M_{V}=-18.8, U-V=1.45,
  V-K=3.13$ in model C.  (d) $M_{V}=-19.2, U-V=1.36, V-K=3.14$ in
  model D.}
\label{fig:idcd}
\end{figure}

In Fig.\ref{fig:idcd}, we show the same figures but for the models C
and D with the starburst model sbB.  The final magnitudes and colours
of picked out luminous ellipticals are $M_{V}=-22.2, U-V=1.55,
V-K=3.19$ and $M_{V}=-22.0, U-V=1.62, V-K=3.25$, in the models C and
D, respectively.  Those of faint ellipticals are $M_{V}=-18.8,
U-V=1.45, V-K=3.13$ and $M_{V}=-19.2, U-V=1.36, V-K=3.14$,
respectively.  In all four panels, the evolutions of metallicities of
galaxies in the spiral state are the same as those in the model A and
B because the difference between the models A and B and the models C
and D is only the starburst model.  In contrast, the metallicities of
galaxies in the elliptical state are different, especially in the high
feedback model and in the dwarf ellipticals (upper- and lower-left
panels and lower-right panel in Fig.\ref{fig:idcd}).  Moreover, the
final metallicities of these four galaxies are almost the same,
$\langle Z_{*}\rangle_{M_{*}}\sim 0.043$.  This value is nearly equal
to the value of the yield $y$.  This shows that the starburst
increases the metallicity to the yield, when all cold gas turns to
stars with very short time-scale.  We will explained this increase of
metallicity in the starburst in the next Section.

From the above results, we conclude that the colour of luminous
galaxies reflects the yield strongly because the effect of the
feedback on such luminous galaxies is very weaker than that on faint
galaxies.  On the other hand, in faint ellipticals, their colour
reflects the feedback efficiency, $\beta$
[$=(V/V_{hot})^{-\alpha_{hot}}$, see eq.(\ref{eqn:alphot})], directly,
when the feedback is effective even in the starburst process.  In
other words, when the value of $y$ is fixed, the evolution of stellar
metallicity in faint galaxies is suppressed in the case of high
feedback models, and the metallicity of luminous galaxies is almost
determined by the yield, $y$, because the feedback effect becomes
negligible in such luminous galaxies.  Therefore, a suppression of
growth of stellar metallicity is required in small galaxies for making
the slope of CMRs.  The relationship between the metallicity and the
feedback is discussed in the next section.

\subsection{Roles of feedback and yield}\label{sec:feedback}
In a chemical evolution of a simple monolithic cloud collapse model,
the evolution of metallicity of cold gas for a constant star formation
time-scale, $\tau_{*}$, is described as
\begin{equation}
Z_{cool}(t)=1-(1-Z_{cool}^{0})\exp\left[-(1-f)\alpha
y\frac{t-t_{s}}{\tau_{*}}\right],
\end{equation}
where $\alpha$ is a locked-up mass fraction, $\alpha=1-R$ ($R$ is the
gas fraction returned by evolved stars), and $Z_{cool}^{0}$ is an
initial metallicity of cold gas at $t=t_{s}$.  The rate of increase of
stellar mass is
\begin{equation}
\dot{M_{*}}(t)=(1-R)\frac{M_{cool}(t)}{\tau_{*}},
\end{equation}
where we use the instantaneous recycling approximation.  From these
two equations, the mass-weighted mean stellar metallicity is obtained
as
\begin{eqnarray}
\langle
\lefteqn{Z_{*}(t)\rangle_{M_{*}}=}\nonumber\\
&&1-F
\frac{1-\exp\left[-(1+\beta-(R-\alpha
y)(1-f))\frac{t-t_{s}}{\tau_{*}}\right]}
{1-\exp\left[-(1+\beta-R(1-f))\frac{t-t_{s}}{\tau_{*}}\right]},
\label{eqn:meanstz}
\end{eqnarray}
where
\begin{equation}
F=\frac{1-Z_{cool}^{0}}{1+\frac{(1-f)\alpha y}{1+\beta-R(1-f)}}.
\label{eqn:meanstz2}
\end{equation}
When the star formation time-scale is enough smaller than the Hubble
time or when the starburst occurs, the final metallicity is
determined by $F$ ($t/\tau_{*}\to\infty$).  The form of $F$ shows that
when the feedback is strong the final mean stellar metallicity becomes
small, $\langle Z_{*}(t/\tau_{*}\to\infty)\rangle_{M_{*}}\to
Z_{cool}^{0}$ as $\beta\to\infty$.  Moreover, increasing $y$ is
corresponding to decreasing the feedback strength $\beta$.  This
effect determines the absolute value of final mean stellar
metallicity, especially for faint galaxies.  On the other hand, when
the feedback is negligible ($\beta\ll 1$), the factor $F$ is
determined by only $y$.  So the effect of $y$ is remarkable in
luminous galaxies rather than in faint galaxies when the starburst
model sbA is adopted, which is the model that the supernova feedback
strength remains even in the starburst.  On the other hand, when the
starburst model sbB is adopted, which is the model that all the cold
gas turns to stars with very short time-scale, the final metallicity of
{\it ellipticals} does not depend on their mass, because $\beta=f=0$
in the starburst.  In this case, we find that the final metallicity
$\langle Z_{*}\rangle_{M_{*}}\sim 0.04$ when $Z_{cool}^{0}\sim
0-0.01$.  This value is almost the same as the final metallicities of
ellipticals in Fig.\ref{fig:idcd}.  Thus the results in the previous
section are mostly explained by eqs.(\ref{eqn:meanstz}) and
(\ref{eqn:meanstz2}).

\section{PARAMETER DEPENDENCE}

In this section, we investigate the parameter dependence of the CMR on
the star formation time-scale and the circular velocity dependence of
the feedback.  In Section \ref{sec:uv}, we also investigate the effect
of the UV background radiation.

\subsection{Star formation time-scale}\label{sec:sft}

In the previous section, the star formation time-scale,
$\tau_{*}^{0}=20$ Gyr, is longer than the Hubble time.  Next we
investigate how the CMR changes in the case of a short star formation
time-scale.  We adopt $\tau_{*}^{0}=2$ Gyr in this section.

\subsubsection{CMR and starburst model}

\begin{figure}
\epsfxsize=8cm
\epsfbox{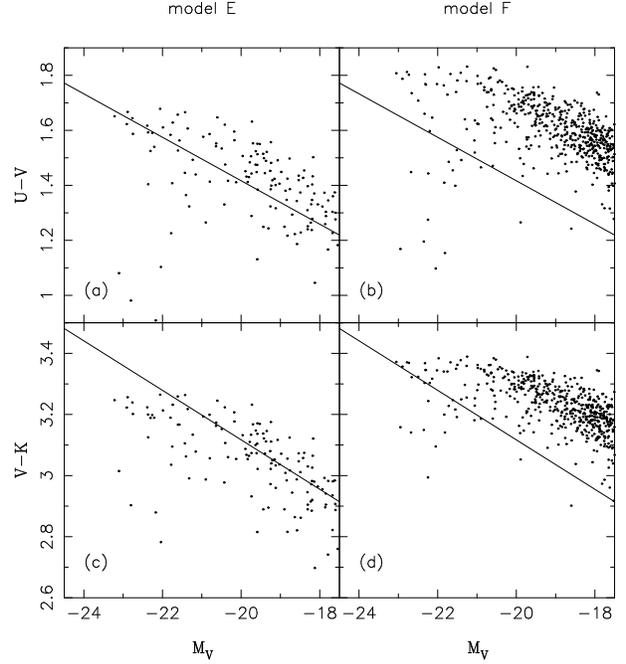}
\caption{Same as Fig.\ref{fig:cmrab}, but for model E and F.
  Both models have a short star formation time-scale, $\tau_{*}^{0}=2$
  Gyr, and the starburst model sbB is adopted.  Model E: $V_{hot}=280$
  km~s$^{-1}$.  Model F: $V_{hot}=100$ km~s$^{-1}$.}
\label{fig:cmref}
\end{figure}

We show the CMRs in the models E and F (Fig.\ref{fig:cmref}).  The
starburst model sbB is adopted in these models.

In the left panels (the model E), the dispersion becomes small
compared to the models A and C and the CMR reproduces the observations
well.  The colours of ellipticals become redder compared to those of
the model A (Fig.\ref{fig:cmrab}) and those of bright ellipticals of
the model C (Fig.\ref{fig:cmrcd}), and become bluer than those of
dwarf ellipticals of the model C.  Because the starburst model of the
model E is the same as that of the model C, these figures show that
the star formation time-scale is one of the essential parameters for
the slope of the CMR when we adopt the starburst model sbB.

In the right panels (the model F), luminous ellipticals are
very red and the CMR is flat at $M_{V}\la-20$.  The colour of the
giant ellipticals nearly corresponds to the colour when $Z_{*}\simeq
0.04$ and the age of galaxies is nearly equal to 10 Gyr.

Moreover, we find that the dependence of CMRs on the starburst model
becomes negligible.  The above results do not change when the
starburst model sbA is adopted (not shown).

These properties are discussed in Section \ref{sec:met}.  In the same
way as Section \ref{sec:result}, we investigate the metallicity and
age of ellipticals next.

\subsubsection{Metallicity and age of ellipticals}\label{sec:met}

\begin{figure}
\epsfxsize=8cm
\epsfbox{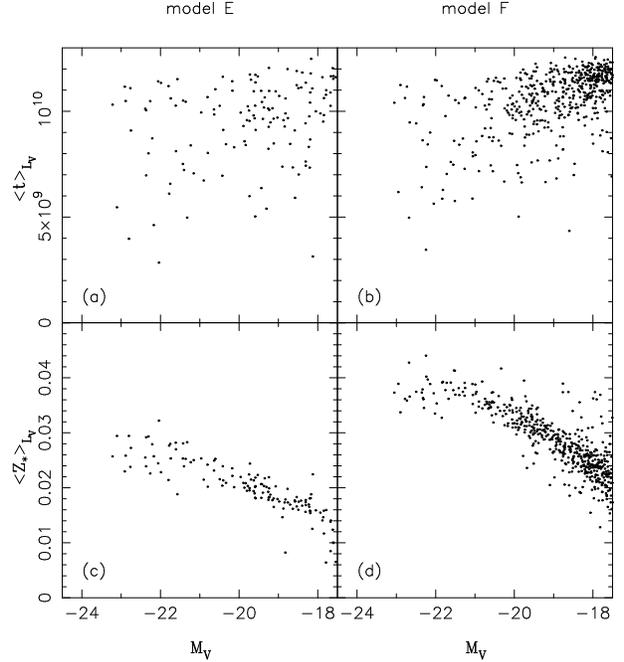}
\caption{Same as Fig.\ref{fig:mtab}, but for model E and F.}
\label{fig:mtef}
\end{figure}

In order to understand the properties mentioned above, we show the
metallicity- and the age-luminosity diagrams in Fig.\ref{fig:mtef}.
Ages of ellipticals become older about 1-2 Gyr than those in the case
of $\tau_{*}^{0}=20$ Gyr on average.  However, this difference
corresponds to less than 0.1 mag in colour.  We cannot explain the
origin of the difference between the CMRs in the models from A to D
and the models E and F only by the age difference.

The metallicity in the model E [Fig.\ref{fig:mtef}(c)] increases in
all range of magnitude compared to the model A
[Fig.\ref{fig:mtab}(c)], and $\langle Z_{*}\rangle_{L_{V}}\sim 0.028$
at $M_{V}\sim -23$ and $\langle Z_{*}\rangle_{L_{V}}\sim 0.018$ at
$M_{V}\sim -19$.  The reason of the metallicity increase is considered
as follows.  In eq.(\ref{eqn:meanstz}) it is only the case of
$t/\tau_{*}\to\infty$ that the final metallicity is determined by only
$F$.  This is correct when the star formation time-scale is quite
shorter than the Hubble time.  However, the value $\tau_{*}^{0}=20$
Gyr is longer than the Hubble time and does not satisfy the above
condition.  The mass-weighted mean stellar metallicity $\langle
Z_{*}(t)\rangle_{M_{*}}$ is a monotonously increasing function about
$t$ in the description in the Section \ref{sec:feedback}.  So the
actual final mean stellar metallicity is lower than $1-F$.  Therefore
when we adopt a short star formation time-scale, mean stellar
metallicities of galaxies increase totally and colours become redder,
compared to the model A.

In contrast to the model A, in the model C with the same starburst
model sbB as the model E, the metallicity is higher than that in the
model E.  The difference between the metallicities in the models C and
E is emphasized at dwarf elliptical scale.  The reason that the
metallicity of dwarf ellipticals is low in the model E is considered
as follows.  In the spiral state, the cold gas turns to stars in very
short time-scale, $\tau_{*}=2(1+z)^{-3/2}$ Gyr.  In this star
formation mode, supernova feedback can reheat the cold gas, and most
stars are formed in the spiral state owing to the short star formation
time-scale.  Therefore, when the major merger occurs, the number of
stars formed through the starburst is small.  Note that when stars are
formed with the feedback, the stellar metallicity becomes lower than
that in the case of the starburst without the feedback, $\beta=f=0$
(see Section \ref{sec:feedback}).  Thus the metallicity becomes small
compared to that in the model C, in which most of stars are formed
during the starburst without the supernova feedback.  Besides the
difference between the feedback models during the starburst does not
affect the metallicity because of the small number of stars formed
during the starburst.

Thus the metallicities of ellipticals in the model E are higher than
those in the model A and lower than those in the model C, and the
results do not depend on the feedback model.  

\subsubsection{Evolution of mean metallicity}

Next we investigate the evolution of metallicities of some sample
galaxies in order to show the time-scale of the metallicity increase
and that most of stars are formed in disc in the models E and F.

\begin{figure}
\epsfxsize=8cm
\epsfbox{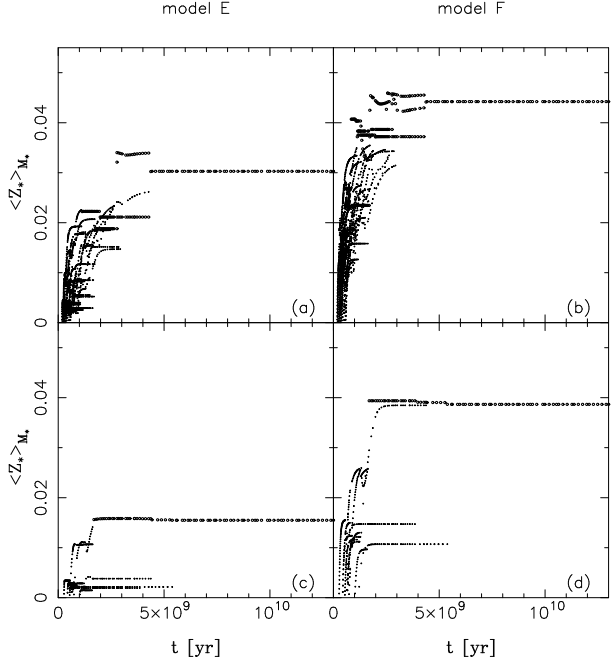}
\caption{Same as Fig.\ref{fig:idab}, but for model E and F.
  (a) $M_{V}=-22.0, U-V=1.61, V-K=3.21$ in model E.  
  (b) $M_{V}=-22.0, U-V=1.77, V-K=3.34$ in model F.  
  (c) $M_{V}=-18.5, U-V=1.44, V-K=3.07$ in model E.  
  (d) $M_{V}=-19.0, U-V=1.60, V-K=3.24$ in model F.}
\label{fig:idef}
\end{figure}

We explicitly show the evolution of the mean metallicity of four
ellipticals in Fig.\ref{fig:idef}.  The magnitudes and colours of the
picked out luminous ellipticals are $M_{V}=-22.0, U-V=1.61, V-K=3.21$
and $M_{V}=-22.0, U-V=1.77, V-K=3.34$ in the models E and F,
respectively.  Those of the faint ellipticals are $M_{V}=-18.5,
U-V=1.44, V-K=3.07$ and $M_{V}=-19.0, U-V=1.60, V-K=3.24$,
respectively.  This figure resembles Fig.\ref{fig:idab} of the models
A and B with the starburst model sbA rather than Fig.\ref{fig:idcd}
of the models C and D with the starburst model sbB.  Note that the
starburst model in the models E and F is the same as that of the
models C and D.  In the models A and B, there are many cases that the
starburst occurs in the middle of the metallicity increase, while in
the end of the metallicity increase in the models E and F.  This shows
that the star formation time-scale is short enough and that most of
stars are formed in disc, where the supernova feedback is effective.
In the models C and D, many stars are formed during the starburst
without the feedback, so that the mean stellar metallicity becomes
very high (see Section \ref{sec:feedback}).  On the other hand, in the
models A and B, the feedback is effective even during the starburst.
So the resulting CMRs of the models E and F resemble those of the
models A and B, while the mean metallicities of the models E and F are
higher than that of the models A and B because of the short star
formation time-scale.

Thus most of stars are formed in disc and the difference between the
CMRs with the starburst models sbA and sbB is negligible when the
star formation time-scale is short enough.

\subsection{Slope of colour--magnitude relation}\label{sec:slope}
In this subsection we investigate how the CMR depends on the parameter
$\alpha_{hot}$.  We adopt $\alpha_{hot}=5.5$ in the models G and H.
This value of $\alpha_{hot}$ has been adopted by Durham group (e.g.,
Cole et al. 1994).  In the model G, $V_{hot}=240$ km~s$^{-1}$.  In
this case, the feedback strength $\beta$ has the same value as that in
the case $\alpha_{hot}=2$ at the Milky Way scale, $V_{c}=220$
km~s$^{-1}$.

\begin{figure}
\epsfxsize=8cm
\epsfbox{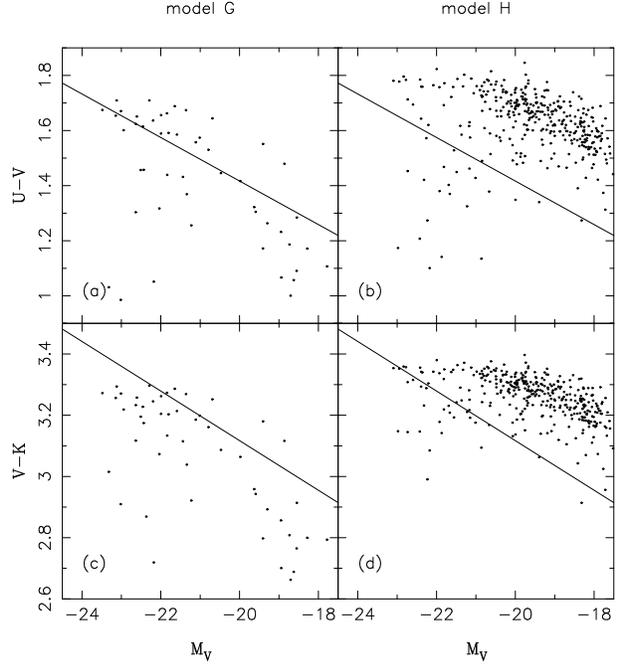}
\caption{Same as Fig.\ref{fig:cmrab}, but for model G and H.  
  Both models have a short star formation time-scale, $\tau_{*}^{0}=2$
  Gyr, and a steep dependence of the feedback on the circular
  velocity.  Model G: $V_{hot}=240$ km~s$^{-1}$.  Model J:
  $V_{hot}=100$ km~s$^{-1}$.}
\label{fig:cmrgh}
\end{figure}

The CMRs in these models are shown in Fig.\ref{fig:cmrgh}.  The
results for the model H with low feedback efficiency (right panels of
Fig.\ref{fig:cmrgh}) are almost the same as those for the model F
(right panels of Fig.\ref{fig:cmref}).  So $\alpha_{hot}$ does not
affect the CMR when the feedback efficiency is low.  On the other
hand, in the model G with high feedback efficiency, the CMR is
different from that in the model E (left panels of
Figs.\ref{fig:cmrgh} and \ref{fig:cmref}, respectively).  The slope of
the CMR is flat at $M_{V}\la -21$ and steeper than that of
observations at $M_{V}\ga -21$.

In the case $\alpha_{hot}=5.5$, the feedback strength $\beta$ changes
drastically at $V_{c}\simeq V_{hot}$.  So galaxies are divided into
two classes at $M_{V}\sim -21$.  Probably this magnitude is given by
$V_{hot}=240$ km~s$^{-1}$ because it is almost the same circular
velocity as that of the Milky Way galaxy, $V_{c}\simeq 220$
km~s$^{-1}$, with $M_{V}\simeq -20.5$.  We find that the metallicity
of the luminous class with $M_{V}\la -21$ is nearly equal to 0.03,
while that of the faint class with $M_{V}\ga -21$ is nearly equal to
0.015.  Moreover, because the star formation in dwarf-scale galaxies
is strongly suppressed, the age of galaxies becomes younger than that
in the case of $\alpha_{hot}=2$.  Therefore the colours of dwarf
ellipticals are bluer than the observations, while those of giant
ellipticals are almost the same colour as the observations.  Thus the
slope of the CMR in the model G seems to be steeper than that of the
observations.

\subsection{Effect of the UV background radiation}\label{sec:uv}

In this subsection, we evaluate the effect of the Ultraviolet (UV)
background radiation on the CMR.  Now we obtain some evidences that
the UV background radiation has existed via the Gunn-Peterson test,
the proximity effect, etc.  The effect of the UV background on galaxy
formation has been evaluated by Nagashima, Gouda \& Sugiura (1999).

The UV photons penetrate galactic gas clouds and ionizing the
envelopes of the clouds.  We assume that absorption of the UV photons
begins at the cooling radius $r_{cool}$ and the UV photons are
perfectly absorbed at a radius $r_{UV}$.  It is assumed that the hot
gas which cannot cool does not affect the UV photons.  The radius
$r_{UV}$ is obtained by the inverse Str{\"o}mgren sphere
approximation,
\begin{eqnarray}
\lefteqn{\int_{r_{UV}}^{r_{cool}}n_{p}(r)n_{e}(r)\alpha^{(2)}(T_{eq})4\pi
r^{2}dr}\nonumber\\
&&=\pi (4\pi r_{cool}^{2})1.5\times 10^{5}J_{-21},
\end{eqnarray}
where $n_{p}$ and $n_{e}$ are the proton and electron number densities
respectively, $\alpha^{(2)}(T_{eq})$ is the recombination coefficient
to all excited levels at a temperature $T_{eq}$, and $J_{-21}$ is the
normalized intensity of the UV radiation, $J\equiv J_{-21}\times
10^{-21}$ erg cm$^{-2}$ s$^{-1}$ sr$^{-1}$ Hz$^{-1}$.  We solve this
equation assuming the isothermal sphere distribution of gas in the
clouds.  The amount of newly cooled gas from hot gas is estimated as
the amount of hot gas within $r_{UV}$.

We have many uncertainties about the evolution of the UV background,
but at low redshift, it is suggested $J_{-21}\propto (1+z)^{4}$ and
$J_{-21}(z=2)\sim 1$ (Pei 1995).  At high redshift, there are too many
uncertainties.  So we approximate the UV background as
$J_{-21}(z=2)=1$, $J_{-21}\propto (1+z)^{\gamma}, \gamma=4$ at
$z\leq2$, $\gamma=-1$ at $2\leq z\leq 5$, and $J_{-21}=0$ at $z>5$.
Nagashima, Gouda \& Sugiura (1999) showed that it is reasonable to
adopt this type evolution of the UV background by comparing the
luminosity function and the colour distribution of galaxies given by
their semi-analytic model with those given by observation.

This effect on luminosity function of galaxies is similar to
increasing feedback efficiency (see Cole et al. 1994; Nagashima et al.
1999).  Therefore, we introduce the effect of the UV background in our
model instead of strengthening the feedback efficiency.

\begin{figure}
\epsfxsize=8cm
\epsfbox{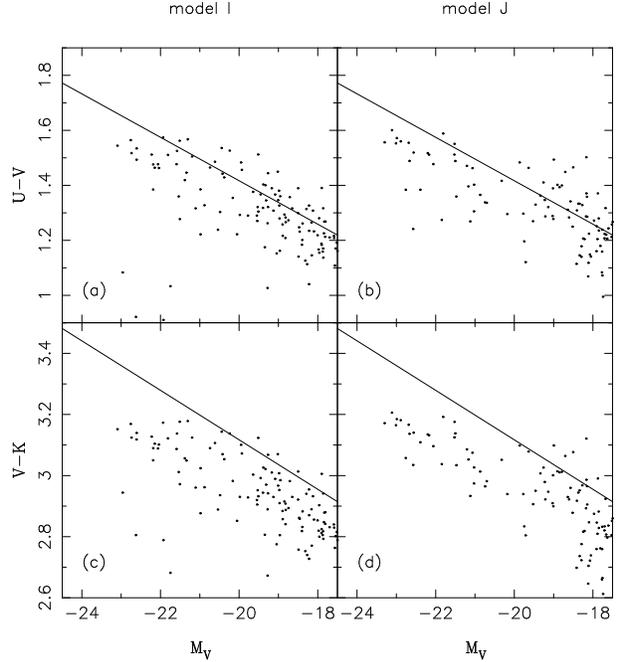}
\caption{Same as Fig.\ref{fig:cmrab}, but for model I and J.  The
  effect of the UV background radiation is considered in model J.
  Both models have a short star formation time-scale, $\tau_{*}^{0}=2$
  Gyr.  Model I: $V_{hot}=400$ km~s$^{-1}$ without the effect of the
  UV background.  Model J: $V_{hot}=280$ km~s$^{-1}$ with the effect
  of the UV background.}
\label{fig:cmrij}
\end{figure}

In Fig.\ref{fig:cmrij}, we show the CMRs in the models I and J, with
short star formation time-scale, $\tau_{*}^{0}=2$ Gyr.  The feedback
strengths are $V_{hot}=400$ and the effect of the UV background is not
introduced in the model I, and $V_{hot}=280$ km~s$^{-1}$ and the
effect of the UV background is included in the model J.  The observed
slopes of the CMR are reproduced well as well as the previous high
feedback models.  

In both models, the slopes of the CMRs remain in the observational
slope, while the colours becomes bluer.  These show that the epoch of
star formation delays by the strong feedback or the UV background
because the star formation occurs in more massive galaxies by the
effects.  In these cases, we also confirmed that the
metallicity-luminosity relations shows almost the same trend as the
previous models.

The virtue of the model J is that the amount of cold gas in `Milky
Way'-like galaxies, which belongs to a halo with circular velocity of
220 km~s$^{-1}$, is $\sim 10^{10}$M$_{\odot}$.  This value is nearly
equal to that in our Galaxy.  On the other hand, the model I
($V_{hot}=400$ km~s$^{-1}$ and $\tau_{*}^{0}=2$ Gyr), the amount of
the cold gas $\sim 2\times10^{9}$M$_{\odot}$ is less than that in our
Galaxy.

Of course, such a {\it degeneracy} will be solved by observing other
quantities, such as luminosity function, cold gas content, and so on,
because the dependences of these quantities on the feedback process,
the star formation time-scale, and the UV background are different.
This will be done in future work.

\section{CONCLUSIONS AND DISCUSSION}

We investigate the physical mechanisms determining the properties of
the CMRs.  We show that the feedback efficiency, the star formation
time-scale, and the yield are the key parameters to determine the
properties of the CMRs.  Moreover, we find that the feedback process
in the starburst strongly affects the slope of the CMR when the star
formation time-scale is long.

The metallicity-luminosity relations, which are directly reflected to
the CMRs, are formed as follows.  At high luminosity, massive galaxies
are dominated, whose feedback strength is weak.  Because the feedback
strength decreases as growing the mass of galaxies, the number of high
metallicity stars increases.  In this case, the mean stellar
metallicity is determined by the yield mainly.  

On the other hand, at low luminosity, we must consider some mechanisms
for suppressing the stellar metallicity.  The following three
parameters determine the mean stellar metallicity of the low
luminosity ellipticals: the star formation time-scale, the feedback
strength, and the intensity of the UV background [see
eq.(\ref{eqn:meanstz})].  The final mean stellar metallicity increases
when the star formation time-scale decreases, because the time-scale of
metal enrichment is given by the star formation time-scale and because
the mean stellar metallicity increases as time passes monotonously in
the case that a monolithic cloud forms stars.  When the feedback and
UV background strength increase, the final mean stellar metallicity
decreases.  The reason for this is that the star formation is
suppressed and then the amount of metals ejected into the gas
decreases.

Therefore one of the solutions to form the metallicity-luminosity
relation is to increase the feedback strength when the star formation
time-scale decreases.  Another solution is to consider the effect of
the UV background radiation.  We find that the effect of increasing
the feedback efficiency is the same effect as increasing the intensity
of the UV background.

The above conclusion suggests that the main reason for reproducing the
CMR in the hierarchical clustering scenario is similar with that in
the wind/collapse model (see Section \ref{sec:feedback}).  That is,
the feedback process of star formation is essential for making the
metallicity-sequence and then, as a result, reproducing the slope of
the CMR.  Then from this point of view, we can say that the
collapse/wind model succeeds in showing the essential process for
reproducing the CMR.  However, in the traditional models such as the
collapse/wind model, the formation epoch of elliptical galaxies is
given by hand.  In real, we should consider the clustering process of
dark haloes, cooling of baryonic gas, merging process of galaxies,
etc.  In fact, the age of elliptical galaxies is not unique.  In the
semi-analytic model, we can pursue such formation processes of
galaxies in the standard scenario based on the hierarchical clustering
which is strongly supported by the present observations of the
large-scale structure of the Universe.  Moreover, in the hierarchical
clustering scenario, processes of suppressing star formation are not
only the supernova feedback but also the UV background radiation.  The
former depends on the star formation time-scale, while the latter
affects all galaxies coeval.  On the other hand, only the wind is the
process of suppressing star formation in the collapse/wind model. In
this paper, we can show how such effects of suppression of star
formation determine the slope of the CMR in the standard cosmological
galaxy formation scenario.

We also find that the feedback process in the starburst is important
to make the CMR when the star formation time-scale is long enough,
$\sim 20$ Gyr.  When all of the cold gas and the metals do not escape
from a {\it galaxy} but used to form stars in the starburst, this
process is the same as the closed-box chemical enrichment model.  If
the star formation time-scale is long, the amount of the cold gas is
large when the starburst occurs.  Therefore the fraction of metal
rich stars becomes large even in dwarf ellipticals.  When the star
formation time-scale is short enough ($\sim 2$ Gyr), the feedback model
in the starburst does not affect the CMR because almost all of the
stars are formed in disc where the feedback is always effective.  Such
a value of star formation time-scale is often used for describing star
formation in early-type disc galaxies (e.g., Arimoto, Yoshii \&
Takahara 1991).

In our model based on the hierarchical clustering scenario, the metals
released to hot gas return to cold gas later.  We must evaluate
whether this metal recycling affects the luminosity-metallicity
relation.
When $t\gg\tau_{*}$, the reheated gas mass and newly formed stellar
mass fractions are
\begin{eqnarray}
\frac{M_{reheat}}{M_{*}+M_{reheat}}&=&\frac{\beta+Rf}{1+\beta-R(1-f)},\\
\frac{M_{*}}{M_{*}+M_{reheat}}&=&\frac{1-R}{1+\beta-R(1-f)}.
\end{eqnarray}
When considering small galaxies, which have a strong feedback
efficiency ($\beta\gg 1$), most of cold gas are transformed to hot
gas.  So it is difficult for such small galaxies to form high
metallicity stars.  It means that the amount of metals formed in the
galaxies is small.  Therefore the hot gas does not evolve chemically.
Thus the effect of metal recycling from hot gas to cold gas is slight.
In order to confirm this, we calculate the CMR by using the model in
which the metals returned from hot gas to cold gas are removed by
hand.  We find that the difference between the results of considering
such metals and removing metals is small, about 0.2 mag lower in $V-K$
in all range of magnitude, and that the difference between the slopes
is negligible.  Therefore the effect of the metal recycling is
negligible with regard to the slope of the CMR.

In this paper, we show the physical mechanisms forming the CMRs and
discuss the possible effects on the CMRs.  We find a kind of
`degeneracy' among the star formation, the feedback, and the UV
background for making the slope of the CMRs.  This degeneracy will be
solved by comparing other statistical properties of galaxies such as
luminosity function and colour distribution in future work.  Moreover,
we should consider the tightness of CMRs.  The tightness may reflect
the dispersion of the mean stellar ages.  However the colour of each
galaxy is not simply estimated through its mean stellar metallicity
and age.  This is probably because a recent star formation by merging
with small galaxies affects the colour.  This effect is more
emphasized in U-V rather than V-K, and the dispersions of the U-V CMRs
are larger than those of the V-K CMRs especially in the case of
$\tau_{*}^{0}=20$ Gyr.  We will investigate the origin of the
tightness minutely in future work.

\section*{ACKNOWLEDGMENTS}    
We wish to thank F. Takahara, Y. Fujita, T. Yano, M. Sasaki, T. Kodama
and N. Arimoto for useful suggestions.
This work was supported in part by Research Fellowships of the Japan
Society for the Promotion of Science for Young Scientists (No. 2265),
and in part by the Grant-in-Aid for Scientific Research (No. 10640229)
from the Ministry of Education, Science, Sports and Culture of Japan.

\bsp
\end{document}